\documentclass[doublecol]{epl2}

\title{Spontaneous separation of two-component Fermi gases in a double-well trap}

\author{Junjun Xu \and Qiang Gu \thanks{Corresponding author. E-mail: \email{qgu@ustb.edu.cn}}}
\shortauthor{Junjun Xu \etal}

\institute{Department of Physics, University of Science and
Technology Beijing, Beijing 100083, China}


\abstract{The two-component Fermi gas in a double-well trap is
studied using the density functional theory and the density profile
of each component is calculated within the Thomas-Fermi
approximation. We show that the two components are spatially
separate in the two wells once the repulsive interaction exceeds the
Stoner point, signaling the occurrence of the ferromagnetic
transition. Therefore, the double-well trap helps to explore
itinerant ferromagnetism in atomic Fermi gases, since the
spontaneous separation can be examined by measuring component
populations in one well.}

\begin{document}

\maketitle

{\bf Introduction.} --Itinerant ferromagnetism has long been one of
the central research topics in condensed matter physics~\cite{mohn}.
Recently, developments of laser trapping and cooling of atomic gases
with multiple components have greatly renewed and broadened the interest
in this area, and the new research interest is mainly expressed in
two aspects.

First, the $\rm ^{87}Rb$ atoms provide an opportunity to study the
itinerant-boson ferromagnetism~\cite{ketterle,Berkeley}. It is the
itinerant-electron ferromagnetism that has been intensively studied
in the context of condensed matter physics. That is a typical Fermi
system. The $\rm ^{87}Rb$ gas comes as the first example of the
ferromagnetic Bose system. Motivated by the experimental
achievement, the ground state properties~\cite{ho,ohmi} and
thermodynamics~\cite{gu1,szirmai} of the Bose ferromagnet have been
studied theoretically. It was shown that the Bose gas is much easier
to exhibit ferromagnetism than the Fermi gas: In the former case,
the ferromagnetic (FM) transition temperature is never below the
Bose-Einstein condensation temperature regardless of the magnitude
of the ferromagnetic coupling~\cite{gu1}, whereas in the latter case
the ferromagnetism can not be present unless the ferromagnetic
coupling exceeds the Stoner point~\cite{mohn}.

Second, cold atomic Fermi gas can be used to simulate mechanism of
the itinerant-fermion ferromagnetism, a long unsolved question in
condensed matter physics. So far, the Stoner model gives us a
qualitative description of itinerant ferromagnetism~\cite{mohn}.
According to this model, the electron system can lower its total
energy by spin polarization when the decrease of interacting energy
is larger than the increase of kinetic energy due to Pauli
principle. Since the interaction between atoms is tunable by
Feshbach resonance \cite{Chin}, people can examine whether such
transition occurs when the repulsive interaction becomes stronger.

Recently an experimental group from MIT claimed their realization of
ferromagnetic phase of Fermi gases in an equally populated mixture
of $\rm^{6}Li$ atoms in the lowest two hyperfine states \cite{Jo}.
They addressed the observation of non-monotonic behavior for
increasing repulsive interactions, which implies the occurrence of
itinerant FM transition through comparing with the Stoner's
ferromagnetic mean-field theory. Previously, Duine and MacDonald had
already investigated features of the FM transition based on
second-order perturbation theory \cite{Duine}. LeBlanc \textit{et
al.} studied observable experimental signatures of the FM transition
theoretically within a local density approximation \cite{LeBlanc}.
More elaborate theoretical studies beyond the mean-field
approximation showed that the FM transition could take place at a
weaker interaction strength \cite{Conduit1}. Moreover, various
related problems, such as textured magnetization \cite{Conduit},
pairing instability \cite{Pekker}, spin fluctuations \cite{Recati},
and population imbalance \cite{Conduit2}, were investigated. On the
other hand, the existence of a FM transition was questioned by Zhai,
who suggested in a phenomenological way that a correlated state
without ferromagnetism could also cause the similar features in the
experiment \cite{Zhai}.

Direct evidence of the FM transition in atomic gases might be the
formation of magnetic domains, or the phase separation of different
spin components. Amoruso \textit{et al.} \cite{Amoruso} and
Salasnich \textit{et al.} \cite{Salasnich} calculated the density
profile of two-component Fermi gases with conserved particle numbers
and demonstrated the spatial symmetry breaking of the system. This
means the formation of a sort of domain-structures. Furthermore,
Sogo \textit{et al.} indicated the similar phenomenon in an open
system with variable particle numbers \cite{Sogo}. The MIT
experiment failed in viewing the domain structure.

It is not an easy job to observe domain-structures in atomic gases
directly. As Jo \textit{et al.} claimed, the lifetime of domains
might be too short and their size might be too small to be observed
\cite{Jo2}. Phase-contrast imaging has been successfully used to
measure magnetization {\it in situ} in $\rm ^{87}Rb$
condensates~\cite{Berkeley}, but {\it in situ} detection for atomic
Fermi gases has not been reported as yet. Some recent theoretical
proposals seek to verify the FM transition indirectly, e.g., by
detecting topological signatures \cite{Berdnikov} and by calculating
the spin drag relaxation rate \cite{Duine2}.

Here we propose an alternative method to detect the spatial symmetry
breaking in the ferromagnetic Fermi gas. This proposal involves
trapping the two-component atomic gas in the double-well (DW)
potential. We show that the two components are spatially separated
and each component prefers staying in one trap if the FM transition
is induced by the strong repulsive interaction. This makes it
convenient to determine the FM transition by measuring atoms in one
well, in comparison to measuring magnetization {\it in situ}.

This paper is organized as follows. In section II, we calculate the
ground state energy of two-component Fermi gases in a DW trap using
the density functional theory. The kinetic term we are using is due
to the so called Thomas-Fermi approximation as used by Sogo
\textit{et al.} \cite{Sogo}. In section III, we show numerical
results for the case with symmetric potential and equal particle
numbers by analyzing the Euler equation. In Section IV, we consider
the asymmetric case by changing the potential depth and the particle
numbers, respectively. The last section shows our conclusions.

{\bf Basic model.} --We now consider the ground state of two-component Fermi gases
trapped in double wells at zero temperature. According to the
density functional theory, the ground state energy of the system can
be written as a functional of the densities:
\begin{eqnarray}
E[\widetilde{n}(r)]=&&\int{d^{3}\widetilde{r}\bigg\{\sum_{\sigma=1,2}
\left[\frac{\hbar^{2}}{2m}\frac{3}{5}(6\pi^{2})^{\frac{2}{3}}\widetilde{n}_{\sigma}^{\frac{5}{3}}
+\widetilde{U}(\widetilde{r})\widetilde{n}_{\sigma}\right]}
\nonumber\\
&& {+g\widetilde{n}_{1}\widetilde{n}_{2}\bigg\}}\label{eq:e1},
\end{eqnarray}
where the first term on the right hand corresponds to the kinetic
energy of the system known as Thomas-Fermi approximation, and it has
the similar form as the energy of uniform Fermi gases. The
Thomas-Fermi approximation is considered a well approach near the
center of the potential trap for large total particle numbers
\cite{Butts}. The second term
$\widetilde{U}(\widetilde{r})\widetilde{n}_{\sigma}$ is the
potential energy of the system within local density approximation,
and
\begin{eqnarray}
\widetilde{U}(\widetilde{r}) =
\frac{1}{2}m\omega^{2}(\widetilde{x}^2 + Ae^{-B\widetilde{x}^2}
+ \widetilde{\rho}^2 ) \label{eq:u2}
\end{eqnarray}
describes the symmetric DW potential. Here $\widetilde{x}$ and
$\widetilde{\rho}$ are the axial and radial coordinate. $A$ and $B$
are the constants related to the shape of the potential well, and
$\sigma=1, 2$ represents the two different component.
$g\widetilde{n}_{1}\widetilde{n}_{2}$ is the interacting energy,
where the strength of coupling is given by
$g=4\pi\hbar^{2}\widetilde{a}/m$ with $\widetilde{a}$ the s-wave
scattering length. The interaction between the same component is
absent as a result of the Pauli exclusion principle.

For simplicity, we transform the integral of equation (\ref{eq:e1})
into cylindrical coordinate and then convert the parameters to
dimensionless forms,
\begin{eqnarray}
\varepsilon=\alpha E,~ x=\beta\widetilde{x},~ \rho=\beta\widetilde{\rho},\nonumber\\
n_\sigma=\zeta\widetilde{n}_\sigma,~ a=\gamma\widetilde{a},\label{eq:e5}
\end{eqnarray}
where
\begin{eqnarray}
\alpha=&&\frac{2^{17}}{3^{7}\pi^{7}\gamma^{8}m\omega^2\xi^{10}},\\
\beta=&&\frac{4}{3\pi\gamma\xi^2},\\
\zeta=&&\frac{128}{9\pi\gamma^3}.\label{eq:e6}
\end{eqnarray}
Here $\xi=\sqrt{\hbar/m\omega}$ is the oscillator length of the trap
and $\gamma$ is a variable relevant to the density profile and the
particle number, which is determined by
\begin{eqnarray}
N_\sigma=\int{\rho dxd\rho n_\sigma}
  =\frac{2^{12}}{3^5\pi^5\xi^6}\frac{1}{\gamma^6}{\widetilde{N}_\sigma},
\end{eqnarray}
where $\widetilde{N}_\sigma$ is the number of each component and
$N_\sigma$ is the reduced particle number.

Then equation ({\ref{eq:e1}) can be simplified as
\begin{eqnarray}
\varepsilon=\int{\rho dxd\rho\bigg\{\sum_{\sigma=1,2}
\left[\frac{3}{5}n_{\sigma}^\frac{5}{3}+U(x,\rho)n_{\sigma}\right]+an_{1}n_{2}\bigg\}}\label{eq:e2},
\end{eqnarray}
where $U(x,\rho)=x^2+ce^{-bx^2}+\rho^2$ is the reduced potential
with $c$, $b$ are constants transformed from parameters $A$ and $B$,
respectively,
\begin{eqnarray}
c=A\beta^2,~~ b=\frac{B}{\beta^2}. \label{eq:A}
\end{eqnarray}

The ground state energy of the system can be obtained by minimize
the energy functional. To ensure the conservation of particle number
of each component, it's necessary to introduce two Lagrange
multiplier $\mu_1$ and $\mu_2$, which just represent the reduced
chemical potentials. The real chemical potential of the system
$\widetilde{\mu}_\sigma$ reads
\begin{eqnarray}
\frac{\mu_\sigma}{\widetilde{\mu}_\sigma}=\frac{2^5}{9\pi^2\gamma^2m\omega^2\xi^4}.
\end{eqnarray}

The ground state energy has to fulfill the variational condition,
$\delta(\varepsilon-\mu_1N_1-\mu_2N_2)/\delta n_\sigma=0$. So we
obtain the following Euler equations,
\begin{eqnarray}
n_1^{\frac{2}{3}}+an_2=M_1(x,\rho),\label{eq:e3}\\
n_2^{\frac{2}{3}}+an_1=M_2(x,\rho),\label{eq:e4}
\end{eqnarray}
where $M_{\sigma=1,2}(x,\rho)=\mu_\sigma-(x^2+ce^{-bx^2}+\rho^2)$.
By solving the Euler equations and keeping the particle number of
each component conserved, we can get the density profile of each
component in the ground state.

{\bf The symmetric case: with equal well depth and population.} --
In the case that the two-component fermions with equal population
are confined in a symmetric DW trap, it is reasonable to assume the
chemical potentials of the two components equal, $\mu_1=\mu_2=\mu$.
For an ideal Fermi system the densities of the two components will
be the same everywhere in the DW trap. In the presence of weak
interactions, the density will not change much. This corresponds to
the symmetric solution of Eqs. (\ref{eq:e3}) and (\ref{eq:e4}). In
addition, Eqs. (\ref{eq:e3}) and (\ref{eq:e4}) have the asymmetric
solution, which means that density profiles of the two components
become different. If the asymmetric solution is stable, the symmetry
of the ground state is spontaneously broken.

\begin{figure}[t]
\includegraphics[width=0.47\textwidth]{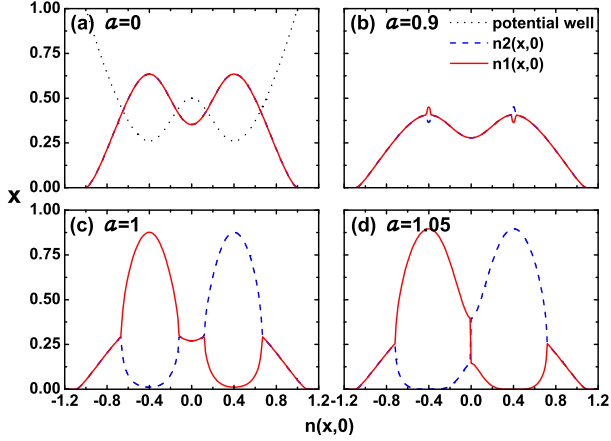}
\caption{Density profile along the axial coordinate $x$ with radial
coordinate $\rho=0$. The reduced interaction strength $a$ from $(a)$
to  $(d)$ is 0, 0.9, 1, 1.05. The dotted-black curve illustrates the
potential well. The solid-red and dashed-blue curves represent
different components. } \label{fig:Fig1}
\end{figure}

\begin{figure}[h]
\includegraphics[width=0.48\textwidth]{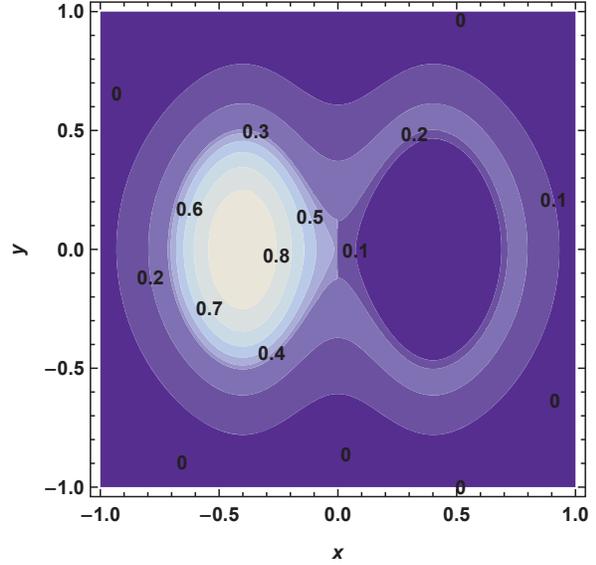}
\caption{Illustration of the density profile $n_1$ along the $x$ and
$y$ coordinate with the interaction parameter $a=1.05$. Numbers in
the figure show the values of the density.} \label{fig:Fig2}
\end{figure}

To minimize the ground state energy functional, one must make sure
that $\partial^2\varepsilon/\partial n^2_\sigma\geq0$. This demands
that the following inequality holds everywhere in the DW trap,
\begin{eqnarray}
n_1n_2\leq\left(\frac{4}{9a^2}\right)^3.\label{eq:e7}
\end{eqnarray}
By analyzing Eqs. (\ref{eq:e3})$\sim$(\ref{eq:e7}), the asymmetric
solution is stable in case that the following condition is
satisfied,
\begin{eqnarray}
M(x,\rho)\in\left[\frac{20}{27a^2},
\frac{1}{a^2}\right].\label{eq:econ}
\end{eqnarray}
There is only the symmetric solution when
$0<M(x,\rho)<\frac{20}{27a^2}$, that is, $n_1=n_2$. Note that there
is no stable solutions when $M(x,\rho)>\frac{1}{a^2}$ according to
Eq. ({\ref{eq:e7}}), which means that the two components are
completely separated, so we have $n_2=0$ and the equation for $n_1$
becomes: $n_1=M(x,\rho)^{\frac{3}{2}}$.

In the whole article we consider a gas composed of $\rm ^6Li$ atoms
with the trap frequency $\omega\approx(170\times2\pi)\mathrm{Hz}$.
First, the total number of two components is supposed to be
$\widetilde{N}=10^6$ ($\widetilde{N}_\sigma=0.5 \times 10^6$) and
the DW potential is $U(x,\rho)=x^2+0.5e^{-10x^2}+\rho^2$. Here the
reduced particle number $N_\sigma$ is normalized to $0.1$ for
convenience and $\gamma\approx 2.57\times10^6m^{-1}$
correspondingly. The potential barrier is defined as
$U_0=\frac{1}{2}m\omega^2A$, then $U_0\approx 15.35h\times {\rm
KHz}$ according to Eqs.~(\ref{eq:A}).

Figure \ref{fig:Fig1} portrays density profiles of the two
components. The spatial width of the confined Fermi gas and the
distance between the two wells are about 2 and 0.8. According to
Eqs. (\ref{eq:e5}), the corresponding physical values are about
119$\mu m$ and 48$\mu m$, respectively. It indicates that the
symmetry-breaking solution becomes stable as long as the interaction
strength is large enough. The two components tend to spatially
separated and each component prefers to gather in one trap. That is
the ferromagnetic state of two-component Fermi gas in double wells.
This state begins to appear at about the interaction strength
$a_c\approx 0.9$, corresponding to the critical scattering length
$\widetilde{a}_c=a_c/\gamma \approx 6600a_0$, where
$a_0=0.529\mathrm{{\AA}}$ is the Bohr radius. As the interaction
strength increases, the two components become more separated. Note
that the components are separated in the inside region of each trap,
but are still equally mixed in the outside region. Figure
\ref{fig:Fig2} illustrates the density profile of one component
along the $x$ and $y$ coordinate with $a=1.05$.

LeBlanc \textit{et al.} estimated that the critical value for $10^6$
$\rm ^6Li$ atoms confined in the single-well with the trap frequency
$\omega\approx(170\times2\pi)\mathrm{Hz}$ is about $6400a_0$
\cite{LeBlanc}. Therefore, trapping atoms in double wells does not
cause much difficulty in generating the FM transition, in comparison
to in the single-well trap. On the contrary, it helps the two
components to be separate in two wells, and thus produces benefits
for the detection.

\begin{figure}[t]
\includegraphics[width=0.47\textwidth]{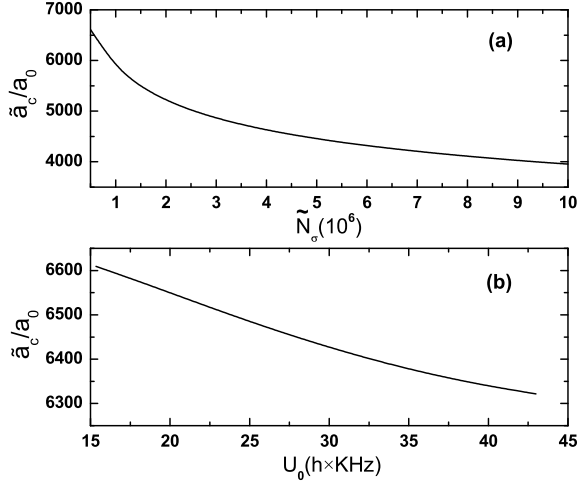}
\caption{The critical scattering length $\widetilde{a}_c/a_0$ as a
function of the particle number $\widetilde{N}_\sigma$ (a) and the
barrier $U_0$ (b). Here $a_0$ is the Bohr radius and $h$
is the Planck constant.
$\widetilde{N}_{\sigma=1,2}=0.5\widetilde{N}$, where $\widetilde{N}$
is the total number.} \label{fig:Fig3}
\end{figure}

Then we discuss the influence of the total particle number and the
potential barrier to the Stoner point. Figure \ref{fig:Fig3}a shows
that the critical scattering length $\widetilde{a}_c$ decreases
apparently with the particle number increasing. Increasing the
barrier $U_0$ but keeping the particle number
$\widetilde{N}_\sigma=0.5\times10^6$ unchanged, the Stoner point is
also lowered, but not that significantly, as shown in
Fig.~\ref{fig:Fig3}b. It is worth noting that there is no
significant change in $k^0_F\widetilde{a}_c$ although
$\widetilde{a}_c$ varies.
$k^0_F=\left[6\pi^2\widetilde{n}_\sigma(0)\right]^\frac{1}{3}$ is
the Fermi wave vector of the interacting gas with
$\widetilde{n}_\sigma(0)$ the density at the center of each well.
Our calculation shows $k^0_F\widetilde{a}_c\approx1.57$, which is
consistent with the mean-field value in the uniform case,
$k^0_F\widetilde{a}_c=\frac{\pi}{2}$.

{\bf The asymmetric case.} --To proceed, we consider the asymmetric
cases caused either by the population imbalance, or by the
difference of well depth. The density profile can still be obtained
by numerically solving Eqs. (\ref{eq:e3}) and (\ref{eq:e4}), with
the particle number of each component being conserved. The problem
of fermion mixtures in a single well with population imbalance has
already been investigated \cite{Conduit2}.

Figure \ref{fig:Fig4} shows the results for the unequal-population
gas in the symmetric DW trap. The ratio of total numbers $N_1:N_2$
is $2:1$. Similar to the equal population case, the two components
distribute symmetrically in the two wells when the interaction
strength is relatively weak. As the interaction grows stronger, this
symmetric distribution becomes unstable. The two components tend to
repel each other and each component begins to occupy the center of
one potential well, signaling the occurrence of the FM transition.

\begin{figure}[t]
\includegraphics[width=0.47\textwidth]{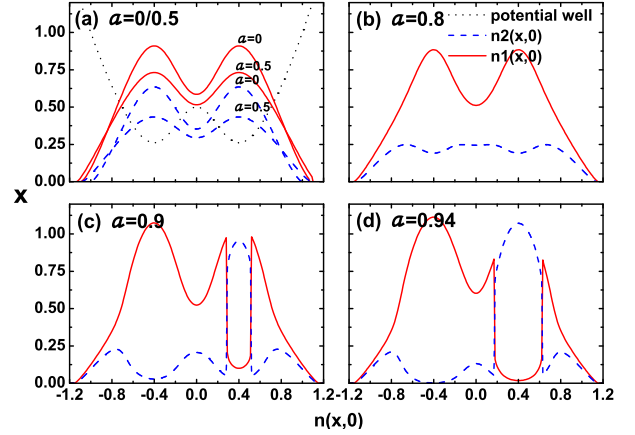}
\caption{Density profile along the axial coordinate $x$ with radial
coordinate $\rho=0$ with unequal particle numbers $N_1=2N_2=0.2$.
The reduced interaction strength $a$ from $(a)$ to  $(d)$ is 0/0.5,
0.8, 0.9, 0.94. The dotted-black curve illustrates the potential
well. The solid-red and dashed-blue curves represent different
components. } \label{fig:Fig4}
\end{figure}

Now look at a two-component Fermi gas with equal population
$N_1=N_2=0.1$ in an asymmetric DW trap. The trapping potential is
chosen as $U(x,\rho)=x^2+0.5e^{-10x^2}+0.5sin(0.5x)+\rho^2$. The
third term in the right hand makes the trap asymmetric in the
$x$-direction. As indicated in Fig.~\ref{fig:Fig5}, the density of
each component equals anywhere in the double trap in the weak
interaction case, although they distribute asymmetric. As the
interaction strength grows, the spatial separation of the two
components occurs first in the deeper well, then in both wells. This
phenomenon is easy to understand since the deeper well traps more
atoms with larger density, so the interacting energy density is
higher than that in the shallower well.

\begin{figure}[t]
\includegraphics[width=0.47\textwidth]{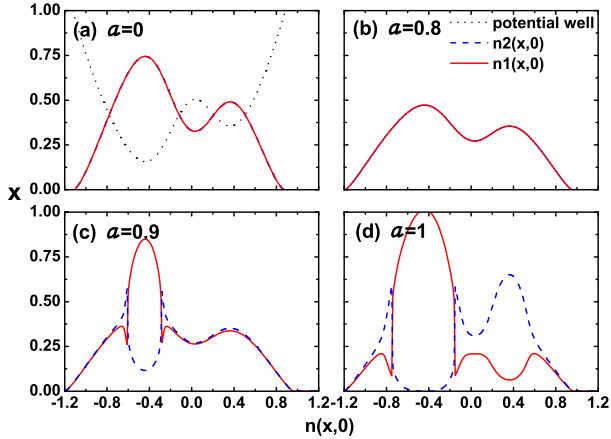}
\caption{Density profile along the axial coordinate $x$ with radial
coordinate $\rho=0$ in the asymmetric double-well potential
$U(x,\rho)=x^2+0.5e^{-10x^2}+0.5sin(0.5x)+\rho^2$. The reduced
interaction strength $a$ from $(a)$ to  $(d)$ is 0, 0.8, 0.9, 1. The
dotted-black curve illustrates the potential well. The solid-red and
dashed-blue curves represent different components. }
\label{fig:Fig5}
\end{figure}

\begin{figure}[h]
\includegraphics[width=0.45\textwidth]{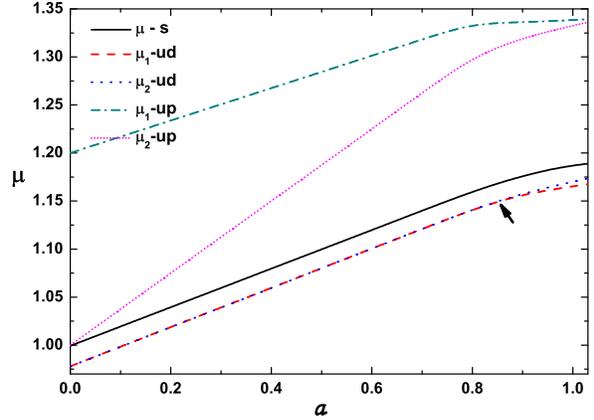}
\caption{The reduced chemical potential $\mu$ versus the interaction
parameter $a$ for different cases. Here $s$ is short for symmetric
case, while $ud$ and $up$ correspond to unequal depth of potential
well and unequal population. Note that $\mu=\mu_1=\mu_2$ for $s$
case and the black arrow shows the transition point in the $ud$
case.} \label{fig:Fig6}
\end{figure}

At last, we illustrate the reduced chemical potential $\mu$ for both
the symmetric and asymmetric cases in Fig.~\ref{fig:Fig6}. It can be
seen that the chemical potential monotonously increases with the
interaction strength for all cases. For the symmetric-potential and
equal-population case, the two components have the same chemical
potential, even if they are spatially separated in the ferromagnetic
state. In contrast, for the asymmetric-potential and
equal-population case, chemical potentials of the two components
have the same value at first, and then become different when the two
components start to separate (see the black arrow in
Fig.~\ref{fig:Fig6}). In the case of unequal-population, each
component has different chemical potential. Comparing
Fig.~\ref{fig:Fig4} with Fig.~\ref{fig:Fig5}, it seems that the
density profiles for the two asymmetric cases display some similar
features when the interaction is strong enough, as shown in
Figs.~\ref{fig:Fig4}d and \ref{fig:Fig5}d.

{\bf Conclusion.} --In conclusion, we have studied density profiles
of two-component Fermi gases in double wells using density
functional theory. Both symmetric and asymmetric DW potentials have
been taken into account and the unequal-population Fermi gas in the
symmetric DW trap has also been discussed. For all the cases, the
obtained results indicate that the two components can be spatially
separated if the repulsive interaction becomes strong enough. This
implies that the ferromagnet aligns in the z (population-imbalance)
direction. In the ferromagnetic state, each component tends to
gather in one well and thus its population dominates in this well.
Therefore, the occurrence of the FM transition can be examined by
measuring atomic populations in one well. It is much easier than
detecting magnetic domains {\it in situ}.

Finally, we note that our calculations may not be directly
applicable when the ferromagnetic alignment is in an in-plane
direction. Moreover, the present study can be extended in several
directions. The critical point needs to be calculated more
accurately. Calculations beyond mean-field theory and numerical
simulations reveal a lower $k_Fa\approx0.8-1.1$ in the single
well\cite{Conduit1}. It is natural to expect that it is also the
case in the double well. Recent works also suggested that some
nontrivial effects may play important role in understanding the FM
transition, such as spatially modulated magnetism \cite{Conduit},
pairing effect \cite{Pekker}, and spin fluctuations \cite{Recati}.

\acknowledgments
This work was supported by the National Natural Science Foundation
of China (Grant No. 11074021), the Key Project of the Chinese
Ministry of Education (No.~109011), and the Fundamental Research
Funds for the Central Universities of China. The authors thank
Yajiang Hao and Joachim Brand for helpful discussions.

\end{document}